\documentclass{ieeeaccess}
\usepackage{cite}
\usepackage{amsmath,amssymb,amsfonts}
\usepackage{algorithmic}
\usepackage{graphicx}
\usepackage{textcomp}

\def\BibTeX{{\rm B\kern-.05em{\sc i\kern-.025em b}\kern-.08em
    T\kern-.1667em\lower.7ex\hbox{E}\kern-.125emX}}
\usepackage{tabularx}
\usepackage{multirow} 
\usepackage{xcolor}
\usepackage[hyphens]{url}
\usepackage{color, soul}
\usepackage{mathtools}

\usepackage{float}
\usepackage{needspace}

\usepackage{caption}
\newcolumntype{C}[1]{>{\centering\arraybackslash}p{#1}}
\usepackage{mwe}
\begin{document}
\history{Date of publication xxxx 00, 0000, date of current version xxxx 00, 0000.}
\doi{10.1109/ACCESS.2023.0322000}

\title{A quantitative model of takeover request \\time budget for conditionally automated driving } 
\author{\uppercase{Foghor Tanshi}\authorrefmark{1, 2}, 
\uppercase{Dirk Söffker}\authorrefmark{2}, 
\IEEEmembership{Member, IEEE}}

\address[1]{Federal University of Petroleum Resources, PMB 1221 Effurun, Nigeria (e-mail: tanshi.foghor@fupre.edu.ng)}
\address[2]{Chair of Dynamics and Control, University of Duisburg-Essen, Lotharstrasse 1, 47057 Duisburg, Germany (e-mail: soeffker@uni-due.de)}
\tfootnote{The research reported in this paper is partly supported by the Tertiary Education Trust fund (Tetfund) of the Nigerian Government through an Academic Staff Training and Development (AST\&D) Scholarship received by the first author for her Ph.D. study at the Chair of Dynamics and Control, University of Duisburg-Essen, Germany where the experiments were conducted.}

\markboth
{Tanshi \headeretal: A quantitative model of takeover request time budget for conditionally automated driving}
{Tanshi \headeretal:A quantitative model of takeover request time budget for conditionally automated driving}

\corresp{Corresponding author: Foghor Tanshi (e-mail: tanshi.foghor@fupre.edu.ng).}

\begin{abstract}
 In conditional automation, the automated driving system assumes full control and only issues a takeover request to a human driver to resume driving in critical situations. Previous studies have concluded that the time budget required by drivers to resume driving after a takeover request varies with situations and different takeover variables. However, no comprehensive generalized approaches for estimating in advance the time budget required by drivers to takeover have been provided. In this contribution, fixed (7 s) and variable time budgets  (6 s, 5 s, and 4 s) with and without visual imagery assistance were investigated for suitability in three takeover scenarios using performance measures such as average lateral displacement. The results indicate that 7 s is suitable for two of the studied scenarios based on their characteristics. Using the obtained results and known relations between takeover variables, a mathematical formula for estimating takeover request time budget is proposed. The proposed formula integrates individual stimulus response time, driving experience, scenario specific requirements and allows increased safety for takeover maneuvers. Furthermore, the visual imagery resulted in increased takeover time which invariably increases the time budget. Thus the time demand of the visualized information if applicable (such as visual imagery) should be included in the time budget.
\end{abstract}


\titlepgskip=-21pt

\maketitle

\section{Introduction}
\label{sec:introduction}
\PARstart{A}{utomated}  driving systems (ADS) have been integrated into vehicles for more than three decades to increase safety. These systems are grouped into six levels (L) of driving automation namely level 0 to level 5~\cite{SAE:2018}. Level 0 (L0) integrates active physical safety systems that mitigate environmental influences and complement driving inputs in emergencies e.g. anti-lock braking system (ABS). Level 1 (L1) integrates one or more intelligent assistance technologies that provide warnings, other safety information and or momentary assistance such as stabilizing driving maneuvers while the driver is performing all dynamic driving tasks (DDT). While L2 systems cooperate with the driver to achieve some DDT such as automating steering input while the human driver remains responsible for monitoring the traffic scenario and performing the remaining DDTs. In L3, the ADS is equipped to perform all DDT such as steering and lane changes provided that the human driver will be receptive and resume driving when requested. In higher levels, the ADS autonomously performs all driving functions in a limited operational domain in L4 and an unlimited operational domain in L5.

\begin{figure}[h!]
\centering
\includegraphics[width=1.0\columnwidth]{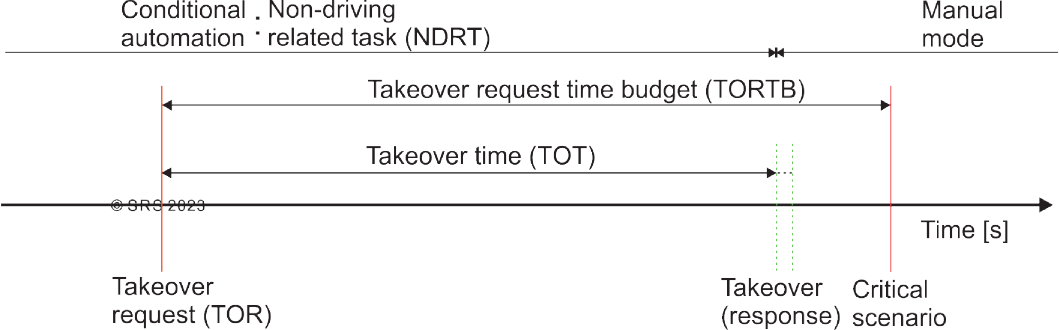}
\caption{Timeline for takeover extended from~\cite{Tanshi:2019c}}~\label{fig:Assisted_TOR_timeline} 
\end{figure}

In L3 which is the focus of this contribution, the human driver becomes momentarily free from driving tasks while the ADS peforms all DDT and may engage in non-driving related tasks (NDRT) such as reading. However, the ADS issues a takeover request (TOR) to the human driver which is also denoted as ``a request to intervene (RTI)" if driving conditions become unsuitable or system limits have been reached~\cite{SAE:2018}. The driver would need to takeover driving within a few seconds and switch back to L0 or L1 as illustrated in Fig.~\ref{fig:Assisted_TOR_timeline}. Ideally, the human driver should takeover successfully within the time budget before the critical situation which may result in a fatal collision. Therefore, to enable a safe and successful transition, the driver requires a sufficient time budget and situation awareness of the traffic context. The situation awareness (SA) is the ``perception of the current situation, comprehension of its meaning, and projection of its status in the near future'' while determining and taking a response~\cite{Endsley:1988}. In addition to SA, other variables associated with conditional takeovers are discussed below.

\subsection{Variables associated with takeovers}

\begin{itemize}
\item Driving scenario variables include the ego vehicle's speed, surrounding traffic density, and the number of lanes. The capacity of drivers to have sufficient SA of these variables in each situation has a significant impact on takeover performance~\cite{Kim:2017}.
\item Takeover time (TOT) is the interval between the TOR and the driver's significant change in steering or pedal input~\cite{Hergeth:2017}. Monitoring of steering and pedal output levels, as well as computer vision-based approaches, are used to detect the significant change in driver input\\~\cite{VOGELPOHL2018,Hergeth:2017}. 
\item Workload also denoted as mental workload is the cognitive and physical demand required by a task~\cite{Hart:1988}.
\item Non-driving related tasks (also known as secondary tasks) are activities that the driver can engage in while the ADS executes the dynamic driving task (DDT), such as reading an e-mail~\cite{VOGELPOHL2018}. The complexity of the NDRT affects the takeover performance due to time and effort required to disengage.
\item Performance variables are measures such as 
mental workload demand of the task, time to collision (TTC) between ego and surrounding vehicles, acceleration and lateral displacement of the ego vehicle etc. An increase in the value of some variables such as workload, lateral displacement, and acceleration indicate poor performance and vice versa. Certain maneuvers such as lane changes and taking exits could briefly result in increased lateral displacement from the lane center. However, if the lateral displacement is consistently high throughout the duration of the maneuver, then a poor performance may be observed irrespective of whether a collision occurs or not. However, each scenario has to be analysed individually to draw an appropriate conclusion. On the other hand, an increase in the value of other variables such as TTC indicate good performance and vice versa. Thus it is always best to measure lateral displacement over the duration of the maneuver. Although, the TOT is widely used to evaluate performance, it is not sufficient because it does not comprehensively reflect the safety of the takeover which may not have been steady~\cite{WU2022106647}. In addition, besides indicating that the driver has regained significant control of the ego vehicle, the TOT does not indicate whether the driver has completed the maneuvering necessary to prevent the critical situation from happening. 
\end{itemize}


\subsection{Takeover request time budget} 

Many studies have drawn various conclusions related TOR but mainly about the variables that affect TOT and TORTB as well as models and predictions of TOT. In~\cite{Kim:2017}, the authors concluded that drivers prefer long TORTBs to takeover correctly. In~\cite{Hergeth:2017} and~\cite{Wang:2019}, the authors concluded that the first takeover experience requires more time. Similarly, the authors of~\cite{BRANDENBURG2020} concluded that takeover time reduces after a previous experience. In~\cite{Bourrelly:2018}, the authors concluded that a long autonomous driving duration increases TOT while the authors of~\cite{Zhenji_LU2020,HUANG2022106534,YangWTL2023} concluded that TORTB affects decision, TOT and overall performance. Shorter TORTB results in shorter TOT and poorer performances. Furthermore, the authors of~\cite{Shuo_LI2021} concluded that elderly people require variably more time. Accordingly in \cite{LI201978}, the authors concluded that older drivers require an average of 0.37 s more reaction time (RT) and takeover time increases with visual warning imagery.


Furthermore, the authors of~\cite{Wang:2019} concluded that TORTB should be varied because drivers tend to be inattentive after successful takeover experiences. On the other hand, the authors of~\cite{VOGELPOHL2018,ShahiniPWZ2023} concluded that drivers require at least 8 s to takeover. Similarly in~\cite{Du:2020}, the authors investigated 4 s and 7 s TORTB and concluded that drivers prefer the later irrespective of the complexity of the situation. Given that complexity varies between different situations, it is unlikely that one TORTB is suitable for all situations. 

Some studies have focused on reducing takeover time as much as possible.
Using a voice assistant to continually give information about traffic increased the chances of a timely takeover by 39\% and male participants performed 1.21 times faster than females~\cite{MAHAJAN2021}. Whereas in~\cite{Forster:2017} and \cite{Schoemig:2018}, guiding drivers using an audiovisual and an augmented reality interfaces respectively during takeover were analyzed. The results indicate an increased SA, reduced TOT and improved overall performance compared to without support in the studied scenarios. However, the authors did not include budgeting of TORTB with respect to the studied scenarios.

Furthermore, \cite{WU2021,LI2021} developed non-generalized regression models and obtained coefficients for the studied scenarios. The differences in individual driving skill and stimulus response time as well as the characteristics and complexity of the scenarios were not included in the estimations of TORTB. Furthermore, in a similar study~\cite{AyoubDYZ:2022} TOT is predicted which is not the time budget required to complete the takeover maneuver. Another study~\cite{TanZ2023} assumes the time budget is sufficient and predicts the TOT for scheduled non-urgent takeovers.


A meta-analysis of 129 studies indicates that shorter TOTs are associated with more urgent situations, not using a handheld device, not performing a visual NDRT, previous takeover experience utilizing auditory or vibrotactile interfaces compared to only visual ones~\cite{ZHANG_joost:2019}. The authors concluded that the aforementioned variables are the determinants of TOT. However, other variables such as driving speed and the scenario complexity were not considered.


In~\cite{Berghoefer:2018} the authors concluded that inconsistent driving environment monitoring results in increased TOT which also varies with individual reaction times. Similarly, in~\cite{Eriksson:2017}, the authors concluded that the TOT of different drivers varies with respect to individual reaction times and levels of driving competence. However, the aforementioned studies did not include any approaches to vary TORTB for different drivers. Therefore, it is necessary to define a method for budgeting variable time for informing drivers of a TOR in different scenarios.




\subsection{Problem statement and aim}

Most studies focus on limiting or defining a general time budget (denoted as TORTB in this contribution) that would be sufficient for all drivers to takeover in all situations~\cite{Kim:2017,Du:2020,VOGELPOHL2018}. In some cases, the authors reported increased SA and performance while reducing takeover time. In other studies, the converse was concluded. However, other studies have concluded that TOT varies between repeated drives and scenarios~\cite{Hergeth:2017,Wang:2019}. Furthermore, if the TORTB is too long, drivers delay response due to reduced SA about the reason for the TOR and may encounter accidents~\cite{Wang:2019}. If the TORTB is too short drivers may have insufficient time to respond which could also result in accidents~\cite{Wang:2019}. 



The aforementioned contradictory results indicate that further study into the effect of the TORTB especially with respect to its estimation is required.  Thus the aim of this contribution is to provide an approach for estimating in detail the time required to takeover in different situations to ensure the driver's safety. The time required by drivers in different scenarios would vary and will not be a fixed time for all situations. In one situation, a long TORTB may be required and a shorter TORTB for another situation \cite{PipkornDT2024}.

In this contribution, it is assumed that the ADS can anticipate critical situations in advance and issue TORs to drivers while giving sufficient time for response. It is also assumed that the ADS can automatically analyze the scenario and provide hazard information to the driver at the time of the TOR.

\subsection{Outline of contribution} 
This contribution focuses on the study of the TORTB required by drivers to successfully takeover in different situations. The outline of this contribution includes an introduction and review of studies related to takeover time and takeover request time budgeting. Next, variables which are included in this contribution for the ADS are then discussed. These variables include various TORTBs for scenarios introduced in previous contributions~\cite{Tanshi:2019c,Tanshi:2022} and this contribution is an extension of the aforementioned. In addition, the previous contributions include different comparisons of driver takeover behavior with respect to scenario complexity levels and secondary tasks using a fixed TORTB of 7 s. Finally, the discussion of results, summary, limitation, and outlook are provided.

\section{Methods}


\subsection{Scenarios and NDRT} 
Several studies preceded this contribution. In \cite{Wang:2019}, three highway takeover scenarios where studied using a time budget of 8 s. In two of the scenarios, the takeover request was as a result of a stationary car in front of the ego vehicle while moving at 70 km/hr and 80 km/hr respectively. The third takeover scenario involved a highway exit while the ego vehicle was moving at 80 km/hr. The results indicate that the drivers delayed response because of the long time budget compared to the complexity of the scenarios. Due to the obtained results, the subsequent study integrated scenarios of less, similar and higher complexity together with a reduced time budget of 7 s detailed in the next paragraph.

In the subsequent contribution, three NDRT complexity levels namely reading, proofreading, and proofreading aloud were studied together with four takeover scenarios~\cite{Tanshi:2019c,Tanshi:2022}. Each of the scenarios have two complexity levels that include highways and country roads as well as zero or more traffic agents. Altogether, eight critical situations were designed in the previous contribution and the TORTB used was 7 s. 

\begin{table}[t]
\centering
\caption{Scenarios and head-up display imagery}~\label{tb:TOR_time_budget_imagery}
\begin{tabular}{p{5cm}p{2.2cm}}\hline
	Scenario description & Warning imagery\\\hline
	\multicolumn{2}{p{7.2cm}}{S1: Stationary front car on the highway (Speed: 130 km/hr)}\\ \includegraphics[width= 4cm, keepaspectratio]{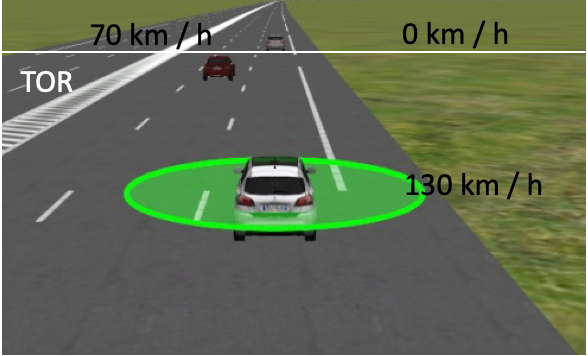} & \includegraphics[width=2cm, keepaspectratio]{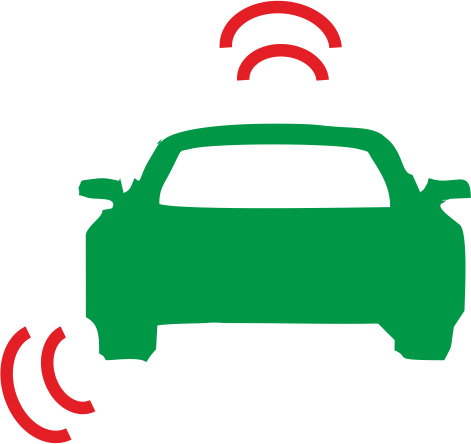} \\
	
	\multicolumn{2}{p{7.2cm}}{S2: Exit the highway (Speed: 50 km/hr)}\\ \includegraphics[width= 4cm, keepaspectratio]{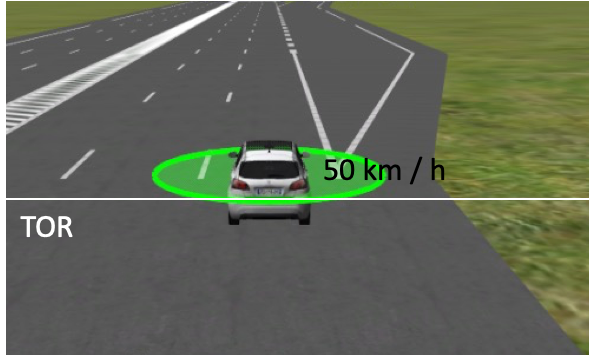} & \includegraphics[width=2cm, keepaspectratio]{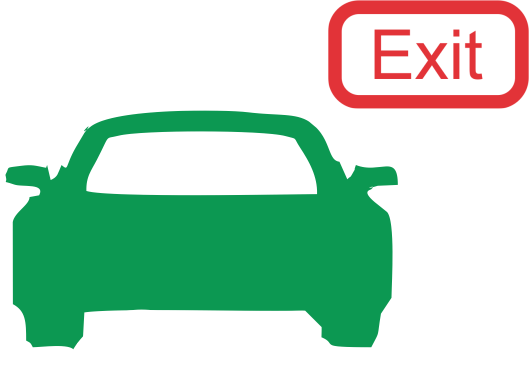} \\
	
	\multicolumn{2}{p{7.2cm}}{S3: Turn right at a country road intersection
		(Speed: 80 km/hr)}\\ \includegraphics[width= 4cm,keepaspectratio]{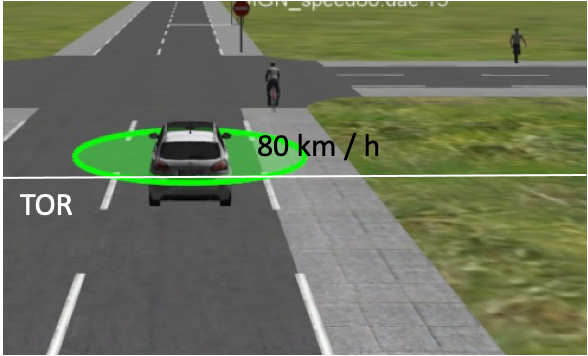} & \includegraphics[ width=2cm, keepaspectratio]{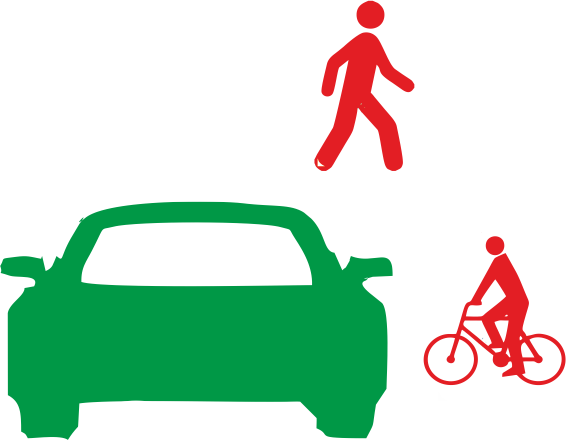} \\\hline 
\end{tabular} 
\end{table}

The results of the 8 scenarios studied in the previous contribution indicated that the second complexity levels of three of the scenarios were more suitable for 7 s as time budget \cite{Tanshi:2022}. In these scenarios, the ego vehicle speed were at least a 100 km/hr with and without interacting traffic agents or had more than one traffic agents together with at least 80 km/hr ego vehicle speed. In addition, the previous results also indicate that the time required by drivers is prioritized in decreasing order of speed, number of traffic agents, and junctions. To further conclude on the time suitability, the two most complex scenarios were chosen together with the least complex scenario while the others were eliminated in this contribution.


The three selected scenarios in this contribution are illustrated in Table~\ref{tb:TOR_time_budget_imagery}. Two of the scenarios (S1 and S3) represent the most complex and one (S2) represents the least complex of the previously designed eight situations. In addition, the three scenarios approximately integrate all the conditions in the previous eight situations. The highway scenarios (S1 and S2) integrates three lanes in two directions and the country road scenario (S3) integrates one lane in two directions. The three NDRTs reading, proofreading, and proofreading aloud were also integrated in this contribution.

\textbf{Stationary car on highway (S1):} In this scenario, the ADS issued a TOR because of a stationary vehicle ahead on the right lane while the ego vehicle was moving at a speed of 130 km/hr. In addition, an approaching vehicle was on the middle lane to the back left side of the ego vehicle. 

\textbf{Exit highway (S2)}: In this scenario, the ADS issued a TOR to the driver to exit the highway while on the right lane (exit lane) and moving at a speed of 50 km/hr.

\textbf{Turn right at country road intersection (S3)}: The ADS issued a TOR to the driver to make a right turn at a four junction country road intersection while moving at a speed of 80 km/hr. A bicyclist and a pedestrian were at the intersection of the right adjourning road where the driver should make the right turn. The pedestrian was crossing the road, while the bicyclist was making a right turn.

\subsection{Design of TORTB Variation}\label{TORTB_var}
\begin{figure}[h!]
	\centering
	\includegraphics[width=0.9\columnwidth]{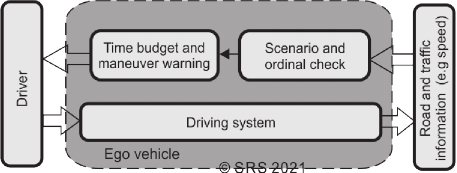}
	\caption{Takeover assistance system (cf.~\cite{Tanshi:2022})}~\label{fig:RF_IR_TOR_assistance_loop} 
\end{figure}

In this contribution various TORTBs were investigated in different takeover scenarios for suitability. The definition of TORTB for the three studied scenarios are based on previous studies~\cite{Hergeth:2017,Wang:2019}. Specifically, existing studies have concluded that budgeting for TOR time should include variation between different scenarios and between first and repeated drives~\cite{Hergeth:2017,Wang:2019}. Specifically, the TOT for the first drive is usually more than that for repeated drives and the TOT for repeated drives are similar. Thus, the TORTB utilized for each scenario in this contribution when a driver experienced it in a first drive was less than when it was in a second or subsequent drive. Based on the characteristic of the scenarios, the assumed order of decreasing complexity is S1, S3 and S2. 

As previously mentioned, the TOT is not sufficient for use as the time budget because it only indicates when the driver has regained significant control of the vehicle without indicating whether the driver has completed the maneuvering required for a particular situation. In addition to previous studies, the design of the time variation is based on conclusions from the preceding contribution~\cite{Tanshi:2022}. The conclusions state that, handsfree NDRTs irrespective of complexity have no significant effect on the time budget and the time budget is the sum of TOT and maneuver completion time. 

\begin{figure}[h]
\centering
\includegraphics[width=1.0\columnwidth]{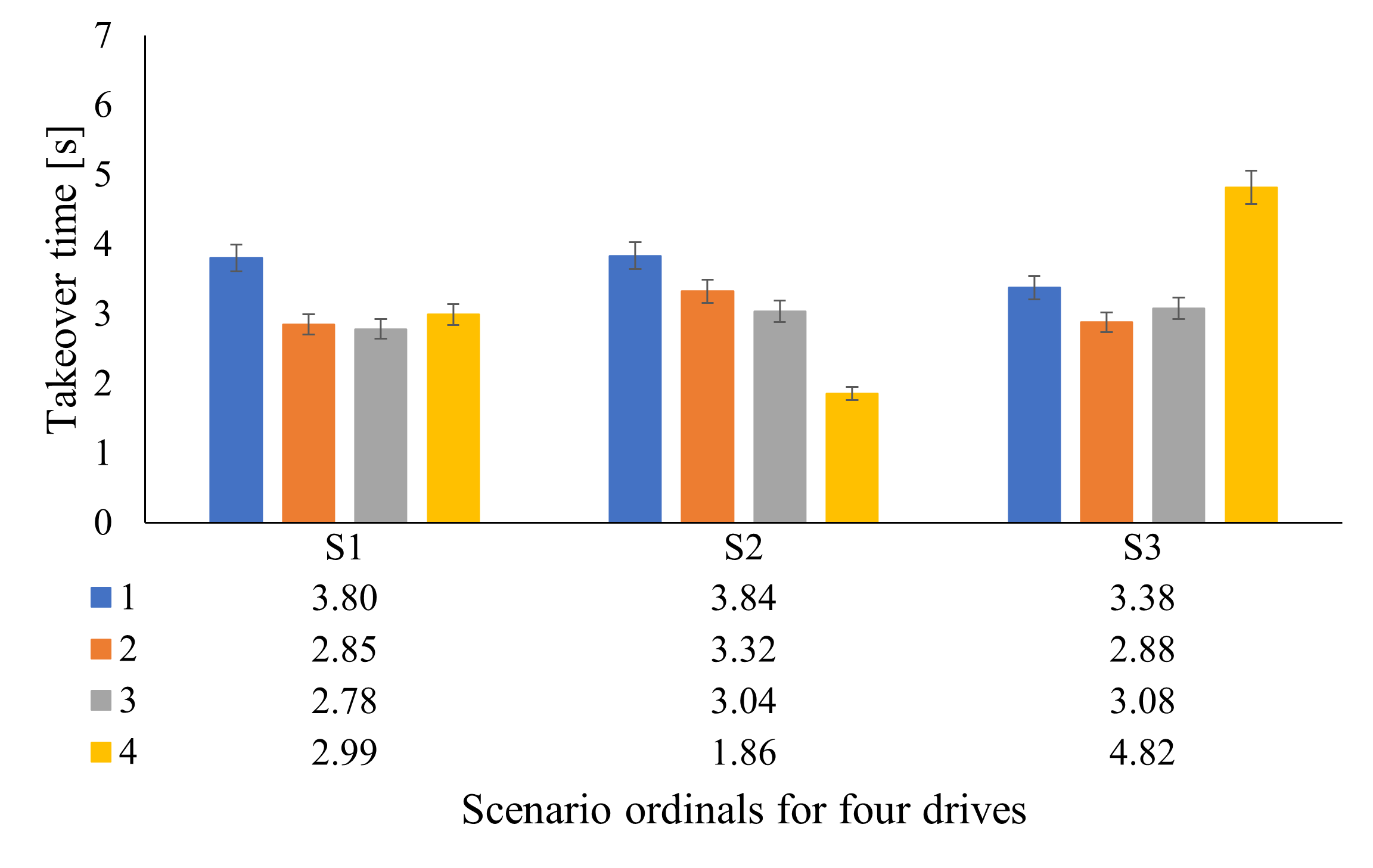}
\caption{Average takeover time for scenario ordinals abridged from~\cite{Tanshi:2022}}
\label{fig:avr_tot_ord_sce_l}
\end{figure}

Based on the aforementioned conclusions, the TOT for the three scenarios in relation to the ordinal (illustrated in Fig.~\ref{fig:avr_tot_ord_sce_l}) from the previous contribution where rounded off (to the nearest integer) and added to the assumed time required to complete the takeover maneuver and avert the critical situation. Given the complexity of the scenarios, the assumed additional time is 2 s and 1 s for the first and second ordinal of S1 and S3 as well as 1 s and 1 s for the first and second ordinal of S2. The additions amount to 6 s (4 s + 2 s) and 4 s (3 s + 1 s) for the first and second ordinal of S1, 5 s (3 s + 2 s) and 4 s (3 s + 1 s) for the first and second ordinal of S3, and 5 s (4 s + 1 s) and 4 s (3 s + 1 s) for the first and second ordinal of S1 as subsequently outlined in Table~\ref{tb:TOR_time_budget}. The numbers were rounded off to the nearest integer and not up because of the need to carefully choose values that were not too high or low which may result in delayed or rushed responses.


\begin{table}[h!]
\caption{Scenario, takeover request time budget, and imagery setting}\label{tb:TOR_time_budget}
\small
\begin{tabular}{p{1.1cm}|p{3.2cm}p{2.5cm}}
	\hline
	Scenario & TORTB & Imagery\\
	\hline
	S1 & G1: 7 s, 7 s, 7 s & G1: No \\
	&  G2 \& G3: 6 s, 4 s, 4 s & G2: No, G3: Yes \\
	S2 & G1: 7 s, 7 s, 7 s  &  G1: No \\
	& G2 \& G3: 5 s, 4 s, 4 s & G2: No, G3: Yes \\
	S3  & G1: 7 s, 7 s, 7 s & G1: No \\
	& G2 \& G3: 5 s, 4 s, 4 s  & G2: No, G3: Yes\\
	\hline
\end{tabular}
\end{table}

\subsection{Experimental participants}\label{participants}
Drivers with different levels of experience categorized as young people were invited to drive in the previously described scenarios \cite{CLARK2017468}. A total of 83 (73 males and 14 females) participants were recruited, among whom 30 have experience with ADS and 31 have experience with driving simulators. The descriptive statistics obtained from the pre-questionnaire include age [Yrs] (mean = 26.4, STD = 3.6, min = 20, max = 35.2), driving experience [Yrs]  (mean = 6.8, STD = 3.8, min = 0.3, max = 17.8), and driving experience [km/wk]  (mean = 238.9, STD = 814.9, min = 5.0, max = 7000).


As their reward, the participants either received 15 EUR or attended a three-hour time management seminar. In accordance with the relevant ethics rules, participants signed a participation consent declaration. The participants were also informed that their participation was voluntary and that they were free to discontinue the experiment if desired.

\subsection{Experimental environment}
\begin{figure}[h!]
\centering
\includegraphics[width=1.0\columnwidth]{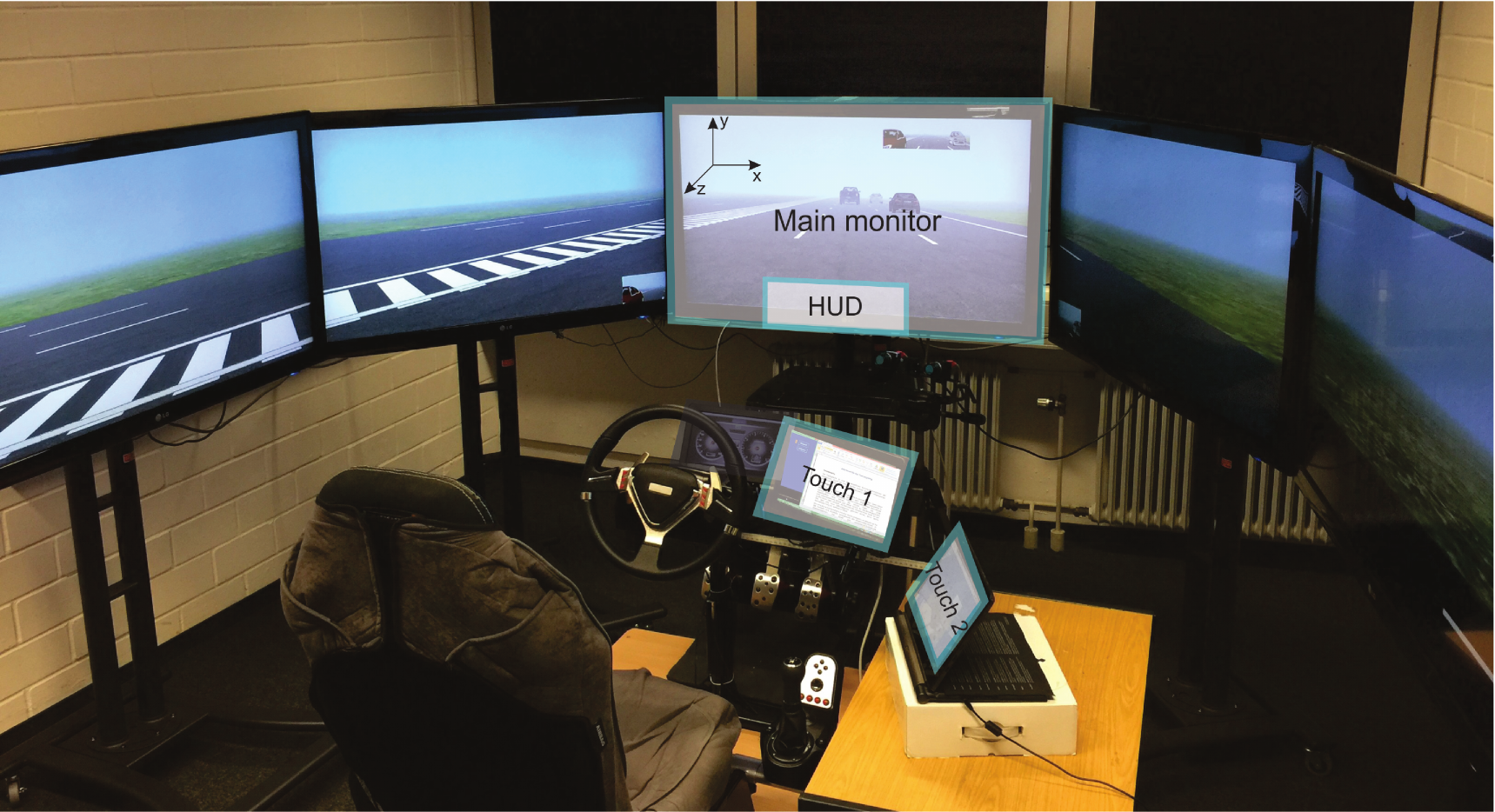}
\caption{Driving simulator lab., Chair of Dynamics and Control, U DuE, Germany}~\label{fig:DS}
\end{figure}

The scenarios were implemented in SCANeR\textsuperscript{TM} studio (a professional driving simulator software by AVSimulation). The data acquiring frequency of SCANeR\textsuperscript{TM} studio is 20 Hz. The driving simulator setup includes five displays that provide 270 \textsuperscript{0} field of view, a fixed-base driver seat, steering wheel, clutch, brake, and accelerator pedals as displayed in Fig.~\ref{fig:DS}. A rear view mirror and two side mirrors were displayed on the appropriate positions of the monitors. A control pad (Touch 1) displays the driving modes on one-fourth of the screen and three-quarter of the screen was used for performing the reading-related NDRTs. In addition, another display (Touch 2) was used by all the participants to complete the questionnaires.

\subsection{Experimental design and procedure}\label{Exp_design}

Furthermore, the participants were divided into three groups namely G1, G2, and G3. Two of the groups G1 and G2 experienced the scenarios without directional warnings while group G3 experienced the same scenarios with the directional warning imagery previously illustrated in Table~\ref{tb:TOR_time_budget_imagery}.  In addition, the TORTB for group G1 participants was held constant at 7 s while those for groups G2 and G3 were varied as previously described in sections~\ref{participants} and \ref{TORTB_var} as well as outlined in Table~\ref{tb:TOR_time_budget}.

\begin{figure}
\centering
\includegraphics[width=0.75\columnwidth]{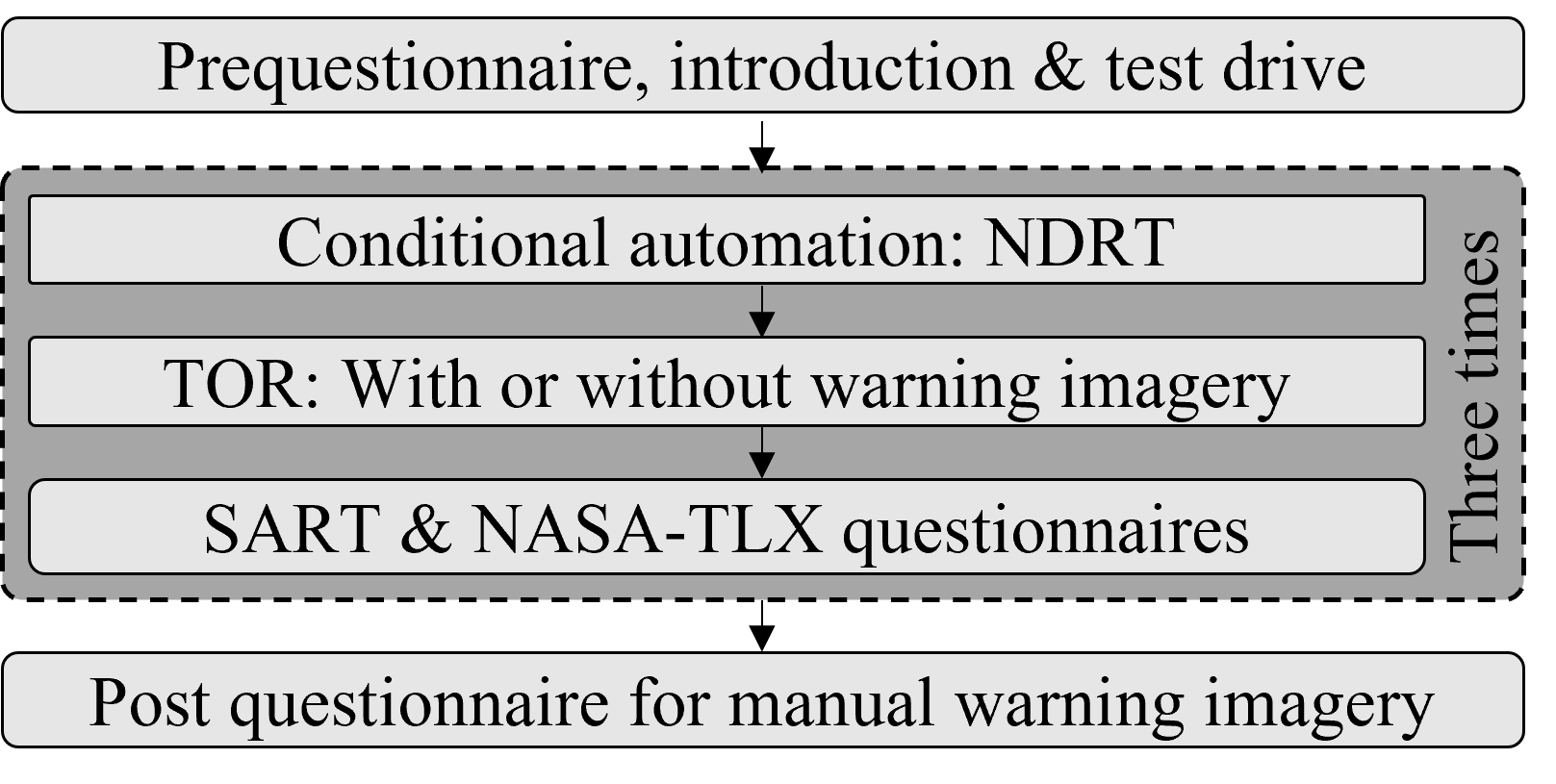}
\caption{Experimental procedure}~\label{fig:exp_procedure}
\end{figure}

Each laboratory appointment lasted approximately two hours and the procedure is summarized in Fig.~\ref{fig:exp_procedure}. First, the participants filled a pre-questionnaire about their driving experience previously detailed in Section~\ref{participants}. Afterwards, the participants received an introduction to the simulator and the procedure as well as the goal of the study for approximately 10 minutes. The visual imagery for the TOR scenarios were not shown to the group G3 participants because takeover situations are meant to be surprising. However, the group G3 participants were informed that the TOR imagery would indicate the position of the collision hazard. The audio sound with which the takeover warnings were given to all groups were played for the participants.

Afterwards, the participants performed a test drive for approximately 10 minutes to adapt to the driving simulator. During the test drive, the participants practiced how to switch between conditionally automated and manual driving modes. Finally, they performed three experimental drive procedures integrating the previously described three takeover scenarios including filling questionnaires as detailed in the next paragraph. The order of experiencing the three takeover scenarios and three NDRTs were randomized between participants to accommodate learning effects and ensured even distribution as first, second, and third drives respectively. Specifically, the participants experienced the scenarios using a 3-by-3 permutation (\textsuperscript{3}P$_{3}$) which results in 6 distinct sequences each that were assigned. Each drive procedure lasted approximately 25 to 30 minutes with a few minutes break in between.

The drivers were first required to drive manually for about 10 to 15 minutes. Afterwards, the participants were asked to switch to conditionally automated mode, during which all the groups performed either one of the three NDRTs. As previously mentioned either of the three NDRTs which are reading related do not have significant effect on the time budget. Afterwards, the TOR was issued to the drivers in the previously stated variable TORTB before each critical situation. To announce the TOR, audio and the previously illustrated directional warning imagery (Table~\ref{tb:TOR_time_budget_imagery}) were utilized for group G3. The directional warnings were used to indicate the reason for the TOR because it is expected to increases drivers' SA~\cite{LI201978,Schoemig:2018}. In addition, the participants from group G1 and G2 received the same TOR audio warning as G3 but without the directional warning imagery. The participants could either switch from conditionally automated to manual mode by touching a button on the control pad or steer and continue driving while applying the pedals. After taking over, the participants filled a situation awareness rating technique (SART)~\cite{Taylor:1990} and raw National Aeronautics and Space Administration task load index (NASA-TLX)~\cite{Hart:1988} questionnaires.


\subsection{Data measures and analysis approach}\label{analysis_approach}

\subsubsection{Data measures}
The measured dependent variables (DVs) included TOT, SA, workload, lateral displacement and maximum acceleration for comparison with the independent variables (IVs) scenarios, ordinal, and group. The ordinal refers to the order of the drive e.g. first, second, or third. As previously mentioned, the ordinal affects TOT such that the first drive requires more time than subsequent ones~\cite{Wang:2019,Hergeth:2017}. The IV ``group" refers to the previously mentioned groups G1, G2, and G3 (Section~\ref{Exp_design}). In other words, a 3 $\times 3 \times 3$ factor statistical analysis approach was employed to study the variation of the between-subject factors (scenario, ordinal, and group) levels. The definition and extraction of measured DVs include

\textbf{Takeover time:} Time from TOR to 5 \% difference in brake or steering input depending on method of takeover used by participant.

\textbf{Situation awareness:} Subjective rating of participant using (SART)~\cite{Taylor:1990}. High ratings indicate increased SA.

\textbf{Workload:} Subjective rating of participant using NA- SA-TLX~\cite{Hart:1988} questionnaires. Increased workload ratings indicate increased task difficulty.

\textbf{Average lateral displacement:} The average lateral displacement (Avg. LD) from the lane center before and after TOR is expressed as  

\begin{equation}
\small
Avg.~LD~[m]=\sum \frac{L.~Displ.~(pre - post)~TOR}{Data~frames~(pre~+~post)~TOR}.
\end{equation}

This is an indication of lateral (steering) control. Increased Avg. LD indicates poor performance~\cite{Kim:2017}. In this contribution, the lateral displacement is compared for a time span before and after TOR which represents a more comprehensive measure of performance and not just at the point of the TOR.

\textbf{Maximum acceleration:} Maximum acceleration value attained between TOR and takeover is expressed as
\begin{equation}
\small
Max.~Acc.~[m~/~s^2] = \max\{Acc.(TOR)~to~Acc.(takoever)\}. 
\end{equation}

This is an indication of forward collision risk. Increased acceleration indicates poor performance\cite{Hergeth:2017}. 

\subsubsection{Analysis approach} 

The analysis for the suitability of the TORTB is from the view that increased TOT does not necessarily indicate more complexity, but performance measures should be considered in the overall conclusions as established in~\cite{Tanshi:2022}. The pattern of variation of the DVs namely TOT, SA, workload, Avg. LD, and max. acc. as well as accident rate and completion of takeover task indicates whether or not the TORTB is suitable. The DVs (TOT, SA, workload, Avg. LD, and max. acc.) were analyzed in relation to the IVs using parametric MANOVA and ANOVA as well as their related outputs that indicate the level of significance. Finally for post-hoc tests, the Bonferroni multi-comparison test was used in relation to the IVs. The subjective measures among the DVs namely SA and workload are used to judge the perceived complexity of the scenarios in comparison to the objective ones. In other words, a decision about the appropriateness of the time budget is objectively analyzed and confirmed or rejected as acceptable based on the subjective perception. This is relevant because the driver's perception that the takeover will fail or succeed will be central to the assessment. The results are subsequently provided in Section~\ref{result_TORTB} and the reasoning is further detailed below.


\begin{itemize}

\item The three groups that the participants are divided into are distinguished by the TORTB and TOR interfaces. The difference between G1 and G2 is the TORTB, between G2 and G3 is the TOR imagery, and between G1 and G3 is the TORTB as well as the TOR imagery that were previously outlined in Table~\ref{tb:TOR_time_budget}. If the TOT in G2 and G3 which have the same TORTB are higher than G1, this indicates that the TORTB for G1 which is higher is more suitable. If the the TOT for G3 is higher than G2, this indicates that the TOR imagery is unsuitable. The applicable result for this reasoning is in Section~\ref{result_TOT}.
\item When scenario complexity increases, TOT and performance often increase. The increasing order of complexity of the studied scenarios are S2, S3, and S1 based on their characteristics. If the TOT for a more complex scenario decreases while performance increases, the TORTB is deemed acceptable and appropriate. On the other hand, TOT may increase with perceived nonurgency in a less complex scenario, indicating that the TORTB for the less complex scenario is too high. In that case, TOT increases and performance decreases in the less complex scenario compared to a more complex scenario. If TOT increases due to nonurgency, the TORTB is deemed unsuitable for that scenario. The applicable result for this reasoning is in Section~\ref{result_TOT}.

\item Similarly for more complex scenarios, SA usually decreases. However, if the TORTB is too long, SA will decrease for a less complex scenario compared to a more complex one, because the driver will delay responding as a result of not immediately perceiving the reason for the TOR. If the driver takes too long to respond, the TOR TORTB is deemed unsuitable. The applicable result for this reasoning is in Section~\ref{result_SA}.

\item Workload frequently increases under time constraints and in more complex scenarios. Workload may decrease due to non-urgency if the TOR TORTB is too long, and vice versa. The applicable result for this reasoning is in Section~\ref{result_WL}.

\item Performance in terms of max. acc. and Avg. LD often decreases with time pressure and for more complex scenarios. However, if decreased performance occurs as a result of a delayed and then sudden response, the TORTB is deemed unsuitable. The applicable results for this reasoning are in Sections~\ref{result_MA} and \ref{result_LD}.
\end{itemize}

Corrupted data and data from four participants who either disobeyed the instructions to perform the NDRTs or did not understand the instructions when they experienced the first takeover were excluded.

%

\section{Experimental results and generalization}


\begin{table*}
\caption{Descriptive statistics of all DVs for scenario and group levels}\label{tb:Descriptives_table_IVs_DVs_2020}
\scriptsize
\begin{tabular}{p{1.35cm}|p{1.4cm}p{1.4cm}p{1.4cm}p{1.4cm}p{1.4cm}p{1.4cm}p{1.4cm}p{1.4cm}p{1.4cm}}\hline
	DVs Mean  & \multicolumn{3}{C{4.2cm}}{S1} & \multicolumn{3}{C{4.2cm}}{S2} & \multicolumn{3}{C{4.2cm}}{S3}\\	
	(STD) & G2 & G3 & G1 & G2 & G3 & G1 & G2 & G3 & G1\\\hline
	TOT [s] & 2.89 & 3.64 & 3.17 & 2.61 & 3.99 & 3.31 & 3.56 & 4.22 & 3.05\\
	& (1.31) & (1.83) & (0.96) & (1.84) & (1.97) & (1.29) & (2.51) & (3.07) & (1.2) \\
	SA  & 12.11 & 10.95 & 16.58 & 24.48 & 23.55 & 24.89 & 15.29 & 15.95 & 20.45 \\
	$[SART]$ & (7.87) & (8.57) & (9.42) & (4.78) & (6.15) & (6) & (7.79) & (8.43) & (9.24) \\
	WL~$[NA-$~& 47.96 & 48.12 & 49.64 & 31.83 & 29.9 & 35.79 & 41.45 & 48.63 & 38.68 \\
	$SA-TLX]$ & (14.35) & (20.79) & (16.81) & (15.85) & (14.69) & (13.32) & (19.94) & (22.78) & (21.06) \\
	Max & 1.16 & 1.68 & 0.08 & 0.45 & 0.61 & 0.17 & 0.15 & 0.16 & 0.12 \\
	Acc.~$[m \ s^2]$ & (1.52) & (1.75) & (0.04) & (0.44) & (0.38) & (0.32) & (0.22) & (0.41) & (0.18) \\
	Avg. LD & 3.52$\times10^{-2}$ & 7.10$\times10^{-2}$ & 1.33$\times10^{-2}$ & 1.05 & 1.16 & 1.39$\times10^{-6}$ & 5.04$\times10^{-8}$ & 1.04$\times10^{-8}$ & 9.56$\times10^{-5}$\\
	$[m]$ & (9.59$\times10^{-2}$) & (1.51$\times10^{-1}$) & (8.27$\times10^{-4}$) & (7.63$\times10^{-1}$) & (7.39$\times10^{-1}$) & (1.14$\times10^{-8}$) & (1.54$\times10^{-7}$) & (9.40$\times10^{-9}$) & (3.83$\times10^{-4}$)\\\hline
\end{tabular}
\end{table*}

\subsection{Experimental results}\label{result_TORTB}
The descriptive statistics of the IVs and DVs are outlined in Table~\ref{tb:Descriptives_table_IVs_DVs_2020}. Two-way Manova indicates significant main and interaction effects of all the IVs on all the DVs combined as outlined in Table~\ref{tb:Manova_table_IVs_DVs_2020}. The main and interaction effects (1-way and 2-way Anova) of the IVs on the individual DVs are outlined in Table~\ref{tb:Anova_table_IVs_DVs_2020}. 

In total, nine accidents occurred in which eight were in S1 (G2 + ordinal 1 = 5, G2 + ordinal 2 = 1, G3 + ordinal 1 = 1, G3 + ordinal 3 = 1) and one in S3 (G2 + ordinal 1 = 1). The increased accident rate in S1 for G2 group indicates an insufficient TORTB. The accident rate decreased in G3 for the same scenarios indicating improved performance. Moreover, more than 60 \% of the participants in G2 and G3 were unable to perform the right turn in S3 compared to less than 5 \% in G1.


\begin{table}[h!]
\centering
\caption{Manova between all IVs and all DVs}~\label{tb:Manova_table_IVs_DVs_2020}
\scriptsize
\begin{tabular}{p{1.95cm}|p{0.65cm}p{0.6cm}p{0.17cm}p{0.7cm}p{1.1cm}p{0.7cm}}\hline
	IV & Wilks'~$\Lambda$ & F & Df & Error~Df & Sig. & Partial~$\eta^{2}$ \\\hline
	Group & 0.277 & 27.405 & 10 & 304 & 4.68$\times10^{-37}$ & 0.474  \\
	Ordinal & 0.347 & 21.198 & 10 & 304 & 8.38$\times10^{-30}$ & 0.411  \\
	Scenario & 0.075 & 80.802 & 10 & 304 & 1.63$\times10^{-79}$ & 0.727  \\
	Ordinal~$\times$~Group & 0.488 & 6.092 & 20 & 505.08 & 1.26$\times10^{-14}$ & 0.164  \\
	Scenario~$\times$~Group & 0.164 & 18.308 & 20 & 505.08 & 9.43$\times10^{-48}$ & 0.364  \\
	Ordinal$\times$~Scenario & 0.202 & 15.642 & 20 & 505.08 & 3.44$\times10^{-41}$ & 0.329  \\
	Ordinal $\times$ Scenario $\times$ Group & 0.304 & 5.227 & 40 & 665.35 & 7.48$\times10^{-21}$ & 0.212  \\\hline
\end{tabular}
\end{table}

\begin{table}[h!]
\centering
\caption{Anova between IVs and DVs}~\label{tb:Anova_table_IVs_DVs_2020}
\scriptsize
\begin{tabular}{p{0.9cm}|p{1.0cm}p{0.45cm}p{0.08cm}p{0.4cm}p{0.73cm}p{0.87cm}p{0.67cm}}\hline
	IVs & DVs & Typ III~SS & Df & MSq & F & Sig. & Partial~$\eta^{2}$ \\\hline
	Group & Tot & 34.5 & 2 & 17.2 & 4.9 & 8.5$\times10^{-3}$ & 0.059 \\
	& MaxAcc & 15.5 & 2 & 7.7 & 16.4 & 3.6$\times10^{-7}$ & 0.17 \\
	& Avg. LD & 5.3 & 2 & 2.7 & 148.9 & 6.7$\times10^{-37}$ & 0.66 \\
	& SA & 441.9 & 2 & 221 & 3.7 & 2.8$\times10^{-2}$ & 0.045 \\
	& Workload & 183.3 & 2 & 91.6 & 0.3 & 7.5$\times10^{-1}$ & 0.0037\\
	Ordinal & Tot & 30.6 & 2 & 15.3 & 4.4 & 0.01.4$\times10^{-2}$ & 0.053 \\
	& Max. Acc. & 6 & 2 & 3 & 6.4 & 2.2$\times10^{-3}$ & 0.075 \\
	& Avg. LD & 4.4 & 2 & 2.2 & 122.5 & 1$\times10^{-32}$ & 6.1$\times10^{-1}$ \\
	& SA & 22.1 & 2 & 11 & 0.2 & 8.3$\times10^{-1}$ & 2.3$\times10^{-3}$ \\
	& Workload & 20.1 & 2 & 10.1 & 3.2$\times10^{-2}$ & 9.7$\times10^{-1}$ & 4.1$\times10^{-4}$ \\
	Scenario & Tot & 8.3 & 2 & 4.2 & 1.2 & 3.1$\times10^{-1}$ & 1.5$\times10^{-2}$\\
	& MaxAcc & 22.6 & 2 & 11.3 & 23.9 & 8.6$\times10^{-10}$ & 2.3$\times10^{-1}$ \\
	& Avg. LD & 19.4 & 2 & 9.7 & 544.5 & 4.4$\times10^{-71}$ & 8.7$\times10^{-1}$ \\
	& SA & 3777.2 & 2 & 1888.6 & 31.3 & 3.8$\times10^{-12}$ & 2.9$\times10^{-1}$ \\
	& Workload & 8336 & 2 & 4168 & 13.1 & 5.3$\times10^{-6}$ & 1.4$\times10^{-1}$ \\
	Ordinal  & Tot & 6.2 & 4 & 1.5 & 0.4 & 7.8$\times10^{-1}$ & 1.1$\times10^{-2}$ \\
	$\times$ & MaxAcc & 18.9 & 4 & 4.7 & 10 & 3.2$\times10^{-7}$ & 2$\times10^{-1}$ \\
	Scenario & Avg. LD & 7.3 & 4 & 1.8 & 103 & 9.7$\times10^{-43}$ & 7.3$\times10^{-1}$ \\
	& SA & 131.3 & 4 & 32.8 & 0.5 & 7$\times10^{-1}$ & 1.4$\times10^{-2}$ \\
	& Workload & 1113.7 & 4 & 278.4 & 0.9 & 4.8$\times10^{-1}$ & 2.2$\times10^{-2}$ \\
	Ordinal & Tot & 10.6 & 4 & 2.6 & 0.8 & 5.6$\times10^{-1}$ & 1.9$\times10^{-2}$ \\
	$\times$ & MaxAcc & 3.6 & 4 & 0.9 & 1.9 & 1.2$\times10^{-1}$ & 4.6$\times10^{-2}$ \\
	Group & Avg. LD & 2.1 & 4 & 0.5 & 29.5 & 2.9$\times10^{-18}$ & 4.3$\times10^{-1}$ \\
	& SA & 441.9 & 4 & 110.5 & 1.8 & 1.3$\times10^{-1}$ & 4.5$\times10^{-2}$ \\
	& Workload & 1429.1 & 4 & 357.3 & 1.1 & 3.5$\times10^{-1}$ & 2.8$\times10^{-2}$ \\
	Scenario & Tot & 9.1 & 4 & 2.3 & 0.6 & 6.3$\times10^{-1}$ & 1.6$\times10^{-2}$ \\
	$\times$ & MaxAcc & 14 & 4 & 3.5 & 7.4 & 1.7$\times10^{-5}$ & 1.6$\times10^{-1}$ \\
	Group & Avg. LD & 9.4 & 4 & 2.3 & 131.4 & 6.9$\times10^{-49}$ & 7.7$\times10^{-1}$ \\
	& SA & 128.8 & 4 & 32.2 & 0.5 & 7.1$\times10^{-1}$ & 1.3$\times10^{-2}$ \\
	& Workload & 897.7 & 4 & 224.4 & 0.7 & 5.9$\times10^{-1}$ & 1.8$\times10^{-2}$ \\
	Ordinal & Tot & 37.7 & 8 & 4.7 & 1.3 & 2.3$\times10^{-1}$ & 6.4$\times10^{-2}$ \\
	$\times$ & MaxAcc & 10.9 & 8 & 1.4 & 2.9 & 5.2$\times10^{-3}$ & 1.3$\times10^{-1}$ \\
	Scenario & Avg. LD & 3.5 & 8 & 0.4 & 24.4 & 5$\times10^{-24}$ & 5.6$\times10^{-1}$ \\
	$\times$ & SA & 336.2 & 8 & 42 & 0.7 & 6.9$\times10^{-1}$ & 3.4$\times10^{-2}$ \\
	Group & Workload & 4608.1 & 8 & 576 & 1.8 & 7.8$\times10^{-2}$ & 8.5$\times10^{-2}$ \\\hline
\end{tabular} 
\end{table}

\begin{figure}
\centering
\includegraphics[width=0.8 \columnwidth]{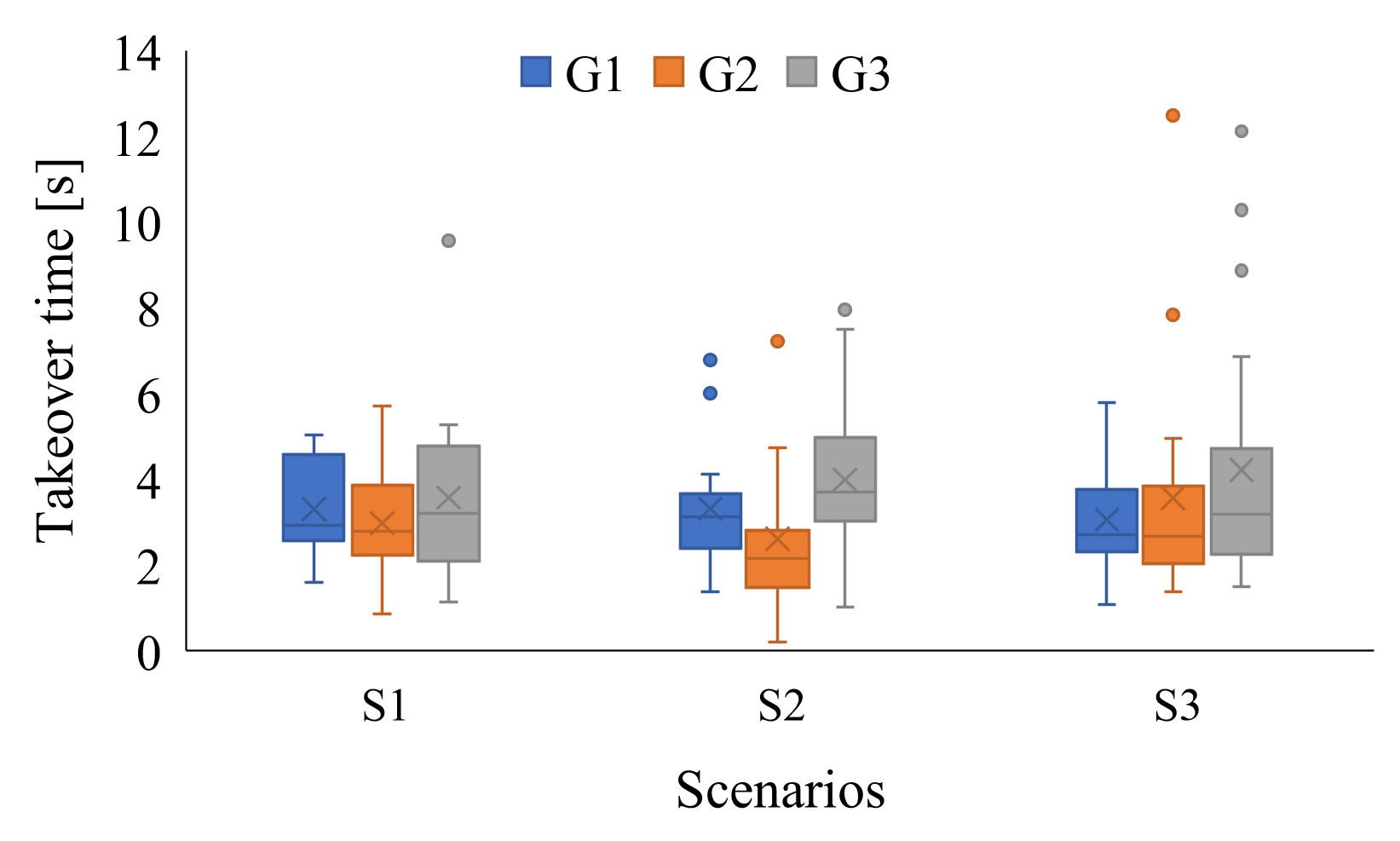}
\caption{Takeover time for scenarios}~\label{fig:ToT_Sce_2020} 
\end{figure}

\begin{figure}
\centering
\includegraphics[width=0.8 \columnwidth]{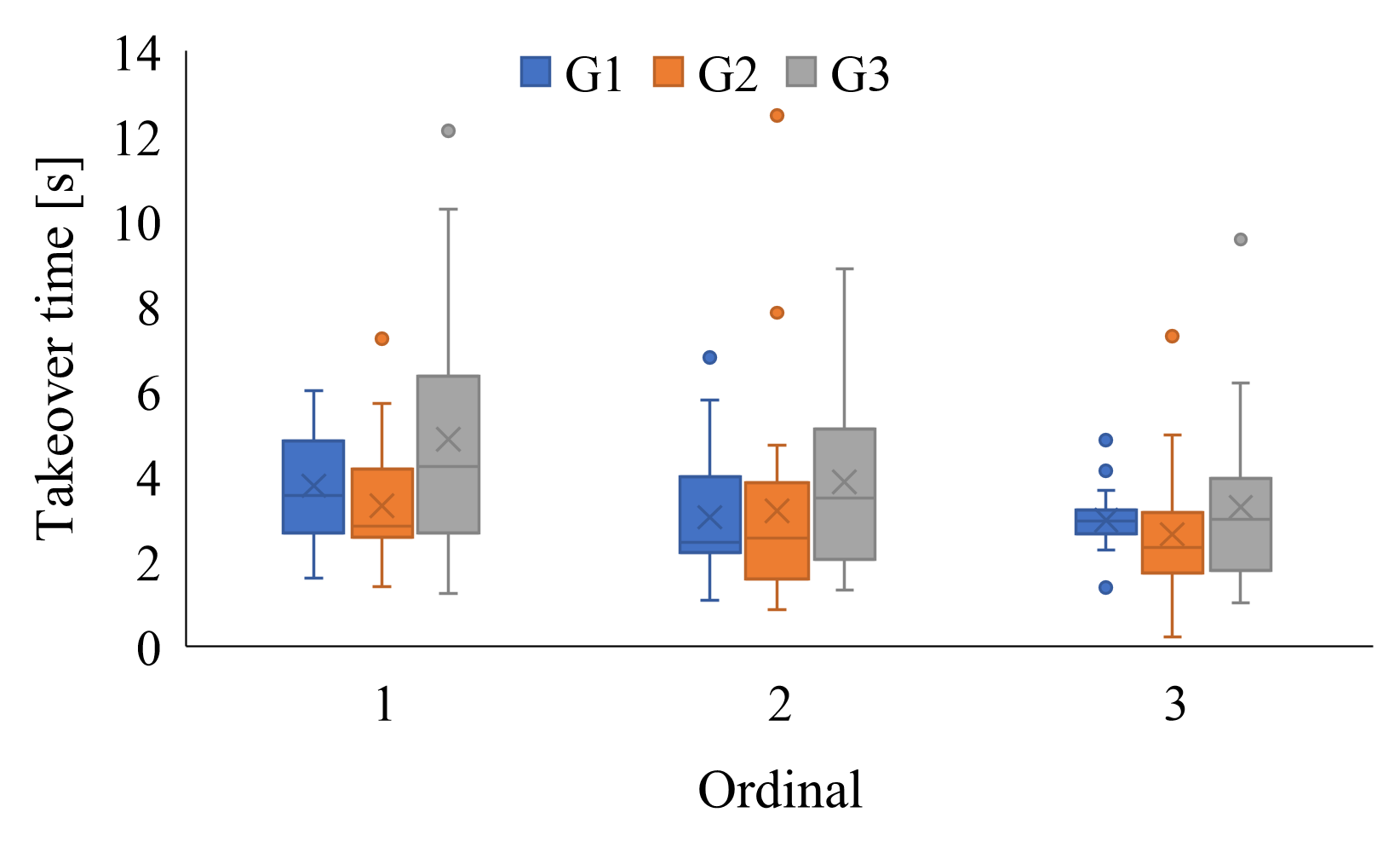}
\caption{Takeover time for ordinal}~\label{fig:ToT_Ord_2020} 
\end{figure}

\subsubsection{Takeover time (Fig.~\ref{fig:ToT_Sce_2020} and Fig.~\ref{fig:ToT_Ord_2020})}\label{result_TOT} No 2-way or 3-way interaction effect between the IVs on TOT was found. No significant main effect of the scenario on TOT exists, indicating that participants always utilized all the available time. A significant main effect of group and ordinal on TOT was found. A post hoc Bonferroni test indicates the following in all three scenarios.

\begin{itemize}
\item[a.] The TOT for groups is such that TOT(G2) $<$ TOT(G3) and TOT(G1) $<$ TOT(G3) indicating that the warning imagery resulted in more visual workload. In other words, the imagery support did not reduce TOT.
\item[b.] The TOT for ordinal is such that TOT(ordinal 1) $>$ TOT(ordinal 2) given that a higher TORTB was assigned to the first drive.  
\end{itemize}

\begin{figure}[h!]
\centering
\includegraphics[width=0.8\columnwidth]{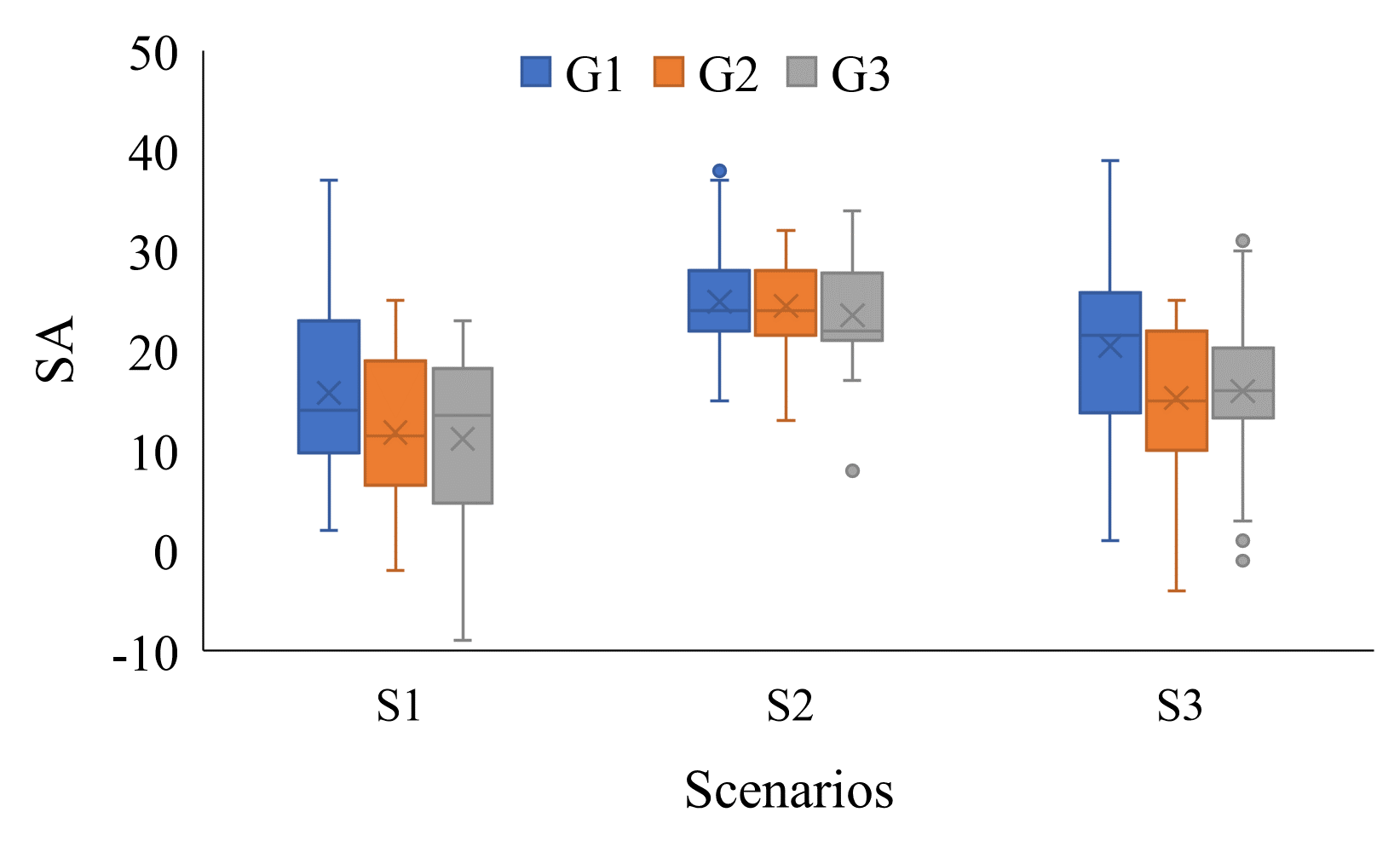}
\caption{Situation awareness for scenarios}\label{fig:SA_Sce_2020}
\end{figure}

\subsubsection{Situation awareness (Fig.~\ref{fig:SA_Sce_2020})}\label{result_SA} No 2-way or 3-way interaction effect between the IVs on SA was found. A significant main effect of group and scenario on SA was found. A post hoc Bonferroni test indicates the following.

\begin{itemize}
\item[a] The SA for scenarios is such that SA(S1) $<$ SA(S3) $<$ SA(S2) which indicates significantly decreased SA in S1 and S3.
\item[b] The SA for groups is such that SA(G3) $<$ SA(G1), indicating that the reduced TORTB of G3 compared to G1 is unsuitable.
\end{itemize}

\begin{figure}
\centering
\includegraphics[width=6.8cm]{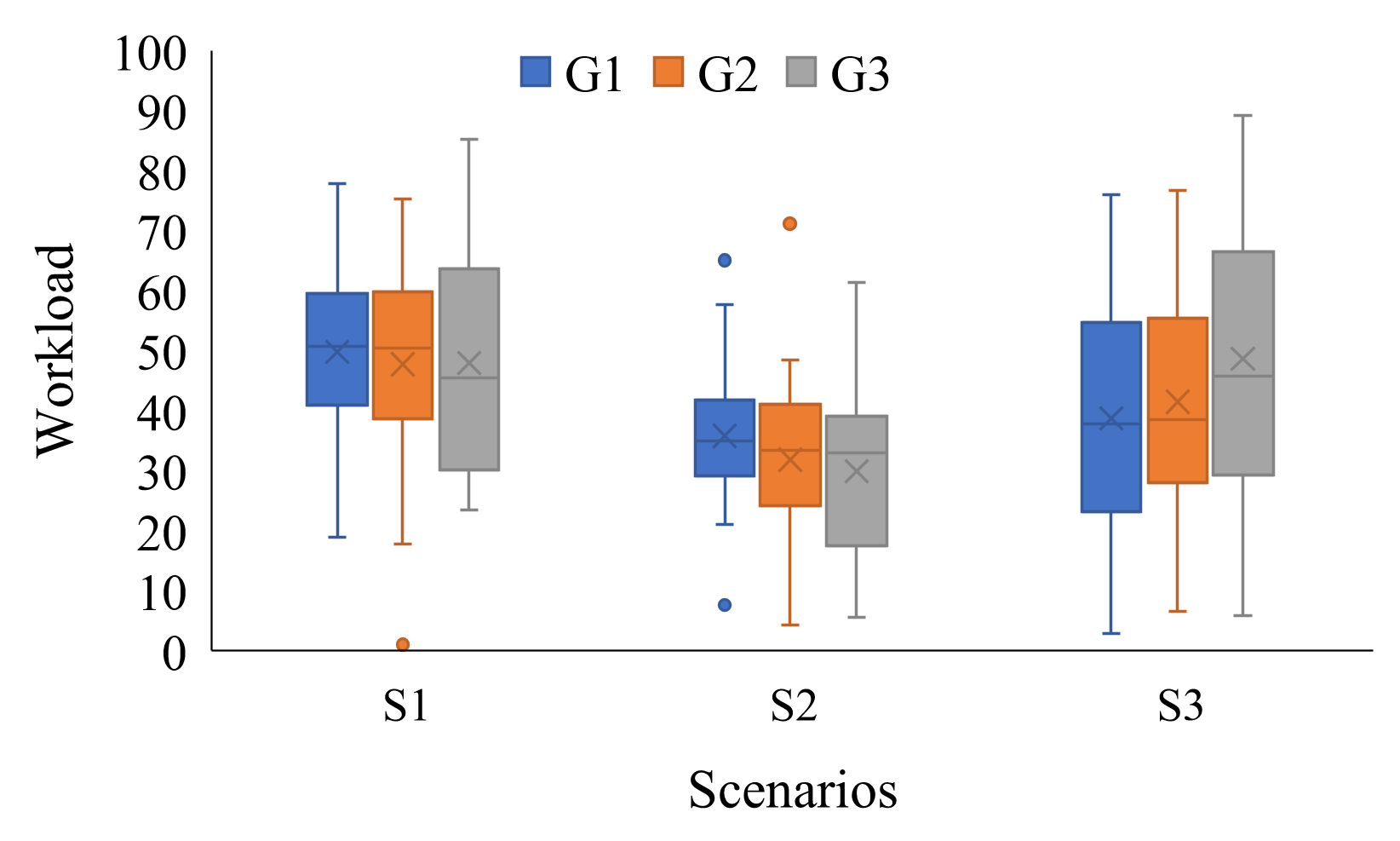}
\caption{Workload for scenarios}\label{fig:Workload_Sce_2020} 
\end{figure}

\subsubsection{Workload (Fig.~\ref{fig:Workload_Sce_2020})}\label{result_WL} No 2-way or 3-way interaction effect between the IVs on workload  exists. In addition, no significant main effects of the ordinal and group on workload exists. A significant main effect of the scenarios on workload was found and post hoc Bonferroni test indicates the following.

\begin{itemize}
\item[a] The workload (WL) for scenarios is such that WL(S1) $>$ WL(S2) and WL(S3) $>$ WL(S2). This indicates a significantly increased and identical workload in S1 and S3 compared to S2 due to the decreased TORTB of G2 and G3.
\item[b] In addition, more than 70 \% of the participants commented that the time was not sufficient to turn right in S3.
\end{itemize}



\begin{figure}
\centering
\includegraphics[width=0.8\columnwidth]{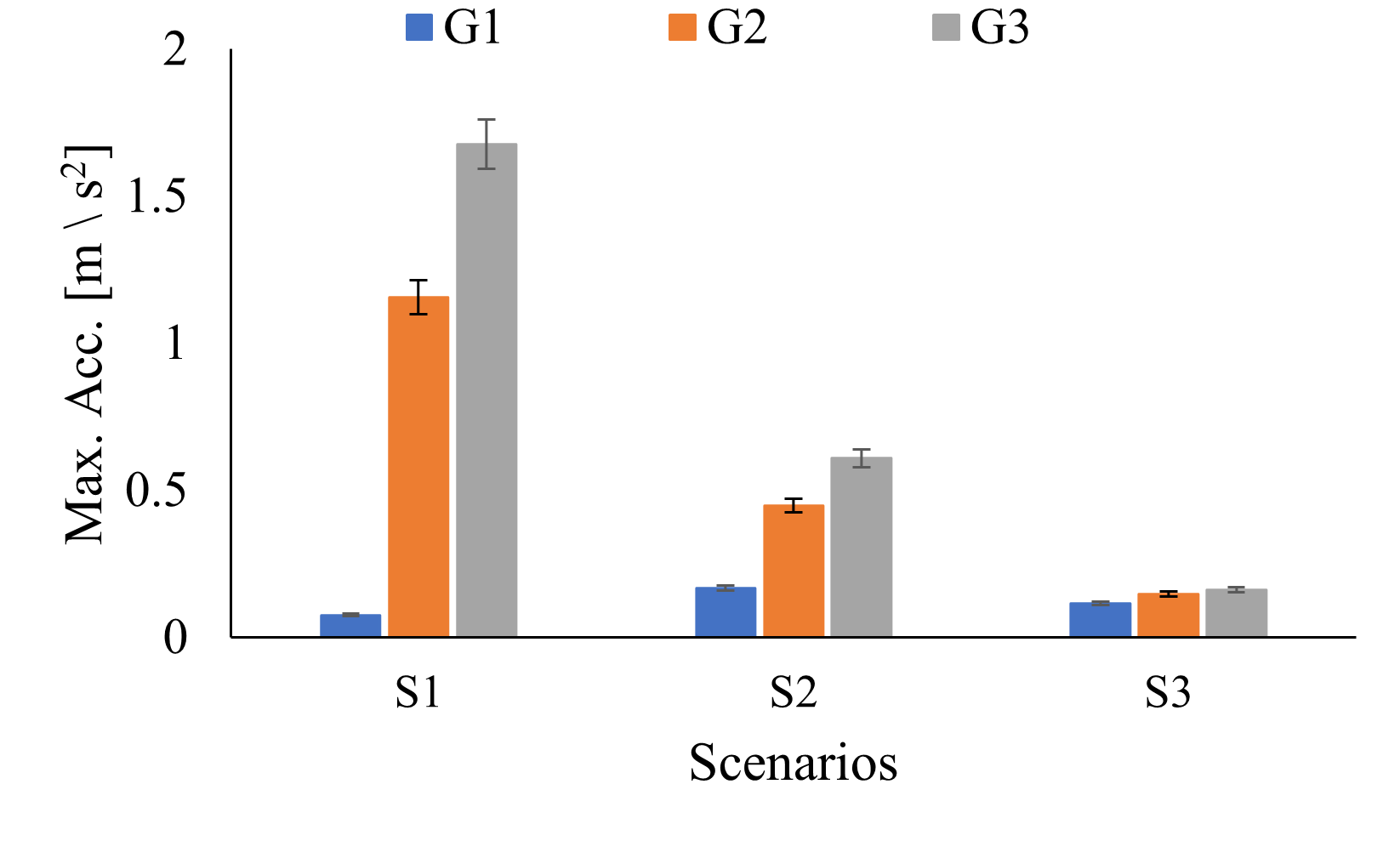}
\caption{Maximum acceleration for scenarios}~\label{fig:Acc_Sce_2020} 
\end{figure}

\subsubsection{Maximum acceleration (Fig.~\ref{fig:Acc_Sce_2020})}\label{result_MA} 
A significant 3-way and 2-way interaction effect between the IVs on max. acc. was found. Furthermore, a significant main effect of ordinal, group, and scenario on max. acc. was also found. A post hoc Bonferroni test indicates the following. 

\begin{itemize}
\item[a.] The max. acc. for scenarios is such that max. acc.(S1) $>$ max. acc.(S2) and max. acc.(S1) $>$ max. acc.(S3). This indicates significantly reduced performance in S1 compared to S2 and S3 due to reduced TORTB. 
\item[b.] The max. acc. for ordinals is such that max. acc.(ordinal 1) $<$ max. acc.(ordinal 2) and max. acc.(ordinal 1) $<$ max. acc.(ordinal 3), indicating better performance in the first drive due to a higher TORTB. 
\item[c.] The max. acc. for groups is such that max. acc.(G2) $>$ max. acc.(G1) and max. acc.(G3) $>$ max. acc.(G1), indicating better performance in group G1 that has a comparably higher TORTB.  

\end{itemize}

\begin{figure}
\centering
\includegraphics[width=0.8 \columnwidth]{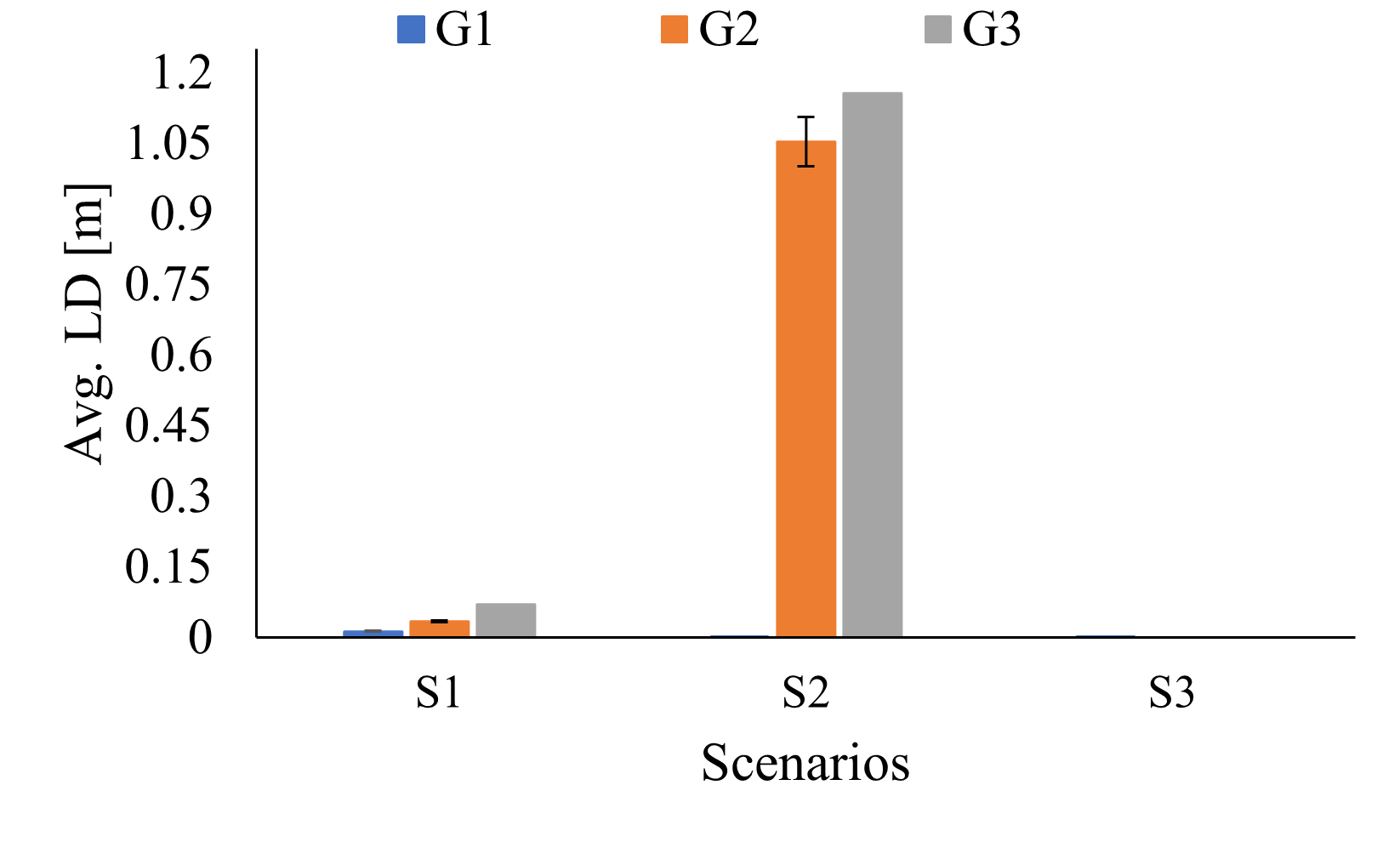}
\caption{Lateral displacement for scenarios}~\label{fig:Lat_displ_Sce_2020} 
\end{figure}

\subsubsection{Lateral displacement (Fig.~\ref{fig:Lat_displ_Sce_2020})}\label{result_LD}

Similar to the max. acc., a significant 2-way and 3-way interaction effect exists between the IVs and Avg. LD. In addition, a significant main effect of the ordinal, group, and scenario on Avg. LD exists.
A post hoc Bonferroni test indicates for Avg. LD that

\begin{itemize}
\item [a.] The Avg. LD for scenarios is such that Avg. LD(S1) $<$ Avg. LD(S2) and Avg. LD(S3) $<$ Avg. LD(S2) which indicates an increased performance in S1 and S3. 
\item[b.] The Avg. LD for ordinals is such that Avg. LD(ordinal 1) $<$ Avg. LD(ordinal 3) $<$ Avg. LD(ordinal 2) indicating better performance in the first drive with a higher TORTB and improved performance in the third drive compared to the second due to learning effects and increased effort. 
\item[c.] The Avg. LD for groups is such that Avg. LD(G2) $>$ Avg. LD(G1) and Avg. LD(G3) $>$ G1 indicating a poorer performance in groups G3 and G2 due to a reduced TORTB.
\end{itemize}


\subsection{Generalization and estimation of TORTB} 
In the different scenarios, the warning imagery and decreased TORTB results in increased TOT in group G3 due to visual demand. Likewise, the SA decreases due to decreased TORTB and increased traffic agents in S1 for G2 and G3 with or without warning imagery. 

Furthermore, the workload with two traffic agents in S1 is identical to S3 in groups G2 and G3 due to a decreased TORTB. In addition, the increased TORTB for G2 and G3 in the first drive of S1 and S3 results in better performance compared to subsequent drives. As previously stated, more than 60 \% of the participants in G2 and G3 were unable to perform the right turn in S3 compared to less than 5 \% in G1. 

The preceding two paragraphs indicate that the 7 s TORTB in G1 is more appropriate for S1 and S3. On the other hand, the TORTB for S2 is inconclusive because besides the TOT, the SA and workload, did not change significantly between the groups. Although the max. acc. and Avg. LD for S2 significantly increased in groups G2 and G3 compared to group G1, this increase may be due to lane change at an exit.

In general, even though the average TOT values measured from a previous contribution~\cite{Tanshi:2022} were adapted and utilized in the same scenarios, the G2 and G3 participants performed worse in S1 and S3 compared to G1. These results indicate that TOT depends on the perception of time sufficiency for success. The results also indicate that the TORTB should be set to the G1 value of 7 s in which the participants performed better and experienced no accidents. Therefore, 7 s was utilized in this contribution as the upper bound TORTB for S1 and S3 that is sufficient for all the participants in the studied age group including the least experienced ones.  

\begin{figure}
\includegraphics[width=0.9 \columnwidth]{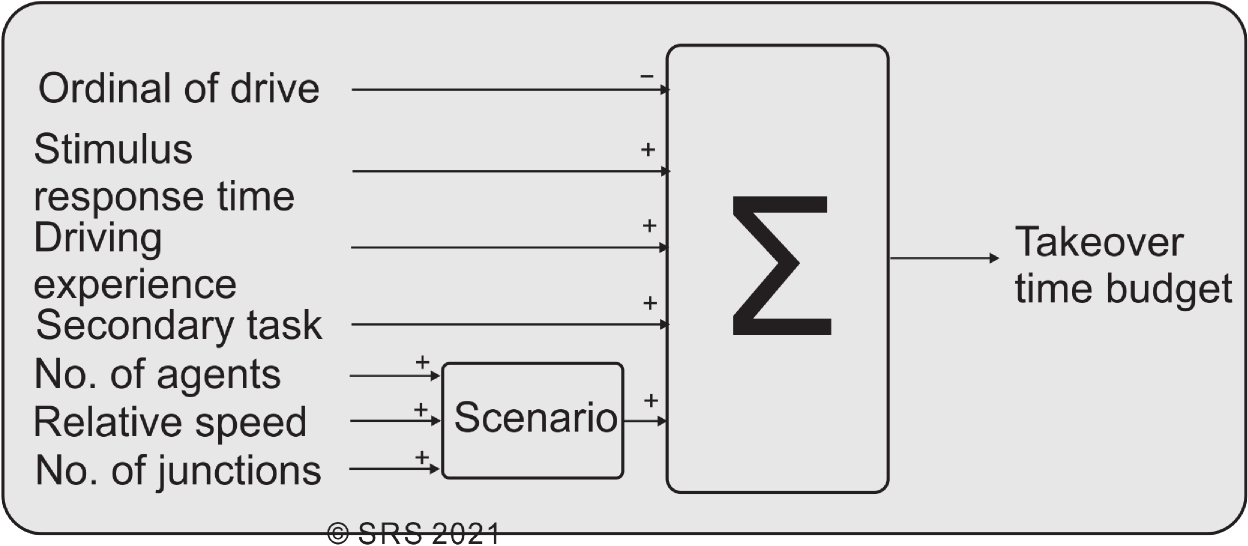}
\caption{Variables for quantitative estimation of takeover request time budget}~\label{fig:TOR_time_budget} 
\end{figure}

Based on the aforementioned results, the effects of the variables for an online TORTB estimation are summarized in Fig.~\ref{fig:TOR_time_budget} as an extension of the scenario and ordinal check step of Fig.~\ref{fig:RF_IR_TOR_assistance_loop}. In addition, the additive nature is based on the known effect of the various variables on the time budget from this contribution and that of other authors. The TORTB estimation follows the sequence of the driver's visual stimulus detection that a TOR has been issued, time demand for disengaging from NDRT, observation of the characteristics of the scenario such as junctions and traffic agents, takeover response due to driving experience, and learning effect due to repeated takeover experience. In accordance with the results obtained in this and previous contributions as well as other related studies, the time budget can be expressed mathematically as
\begin{equation}
\small
TORTB~[s] =
SRT + DEC + SST + NDRTC - OC,\label{eqtib}
\end{equation}

given that
\begin{equation}
\small
SST~[s] = (NOA \times 1.9) + (NOJ \times 0.2) + RSC,\label{eqSST}
\end{equation}

which when substituted in Equation~\ref{eqtib} yields

\begin{equation}\label{eqtib_ex}
\small
\begin{split}
	TORTB~[s] = &
	SRT + DEC + (NOA \times 1.9)~+ (NOJ \\
	&\times 0.2) + RSC + NDRTC - OC
\end{split}
\end{equation}

together with SRT (Stimulus response time), DEC (Driving experience coefficient), SST (Scenario specific time), Number of agents (NOA), NOJ (Number of junctions), RSC (Relative speed coefficient), NDRTC (NDRT coefficient, and OC (ordinal constant) as explained sequentially subsequently. Based on the known suitable value of 7 s for scenarios S1 and S3, the unknowns can be computed and reused in other scenarios.

\textbf{SRT}: Individual SRT affects TOT such that drivers with small SRTs take over faster~\cite{Berghoefer:2018,Eriksson:2017}. Meanwhile, an auditory stimulus requires 140-245 ms, while a visual stimulus requires 180-270 ms and it is affected by intelligence, gender, and physical exercise~\cite{Thompson:1992,Jain:2015}. Accordingly, the visual SRT which is necessary for drivers to perceive that a takeover request has been issued is integrated in the formula. In this contribution, the slowest visual SRT which is equal to 0.3 s rounded off to one decimal place was used to calculate the unknown coefficients. On the other hand, the SRT of drivers can take on any value within the typical range.

\textbf{DEC}:  Individual driving experience affects TOT such that drivers who drive relatively frequently are more skilled and takeover faster~\cite{Berghoefer:2018,Eriksson:2017}. In this contribution, drivers who drive more than 2000 km/wk take over in less than 2 s. To illustrate this, an example of levels generated using the 25$^{th}$, 50$^{th}$, and 75$^{th}$ percentiles of driving experience [km/wk] from the previous contribution is illustrated in Table~\ref{tb:Estimate_SRT_DEC_RSC}\cite{Tanshi:2022}. Furthermore, the DEC is known to be within a typical range of 0.5 s to 1.5 s during active driving and it represents the simultaneous identification, decision, and muscle activation time after observation of the driving scenario~\cite{STEFFAN2013405}. In this contribution, the range is conservatively defined between 1 s to 2 s because takeovers occur suddenly and may differ slightly from active driving because of prior driver disengagement. More so, this conservative definition accommodates the time drivers may need to perceive or feel confident that a takeover maneuver would be successful. Therefore, the least experienced driver's DEC that is used in this contribution to subsequently calculate unknown values is 2 s. The substituted value of DEC is valid for the least experienced driver irrespective of the other example categories in Table~\ref{tb:Estimate_SRT_DEC_RSC}.


\begin{table}[h!]
\centering
\caption{Example estimation of coefficients generated from~\cite{Tanshi:2022,Thompson:1992,STEFFAN2013405})}~\label{tb:Estimate_SRT_DEC_RSC} 
\footnotesize
\begin{tabular}{p{2.15cm}p{1.0cm}p{2.6cm}p{1.0cm}}\hline
RS [km/hr] & RSC [s] & Experience [km/wk] & DEC [s] \\\hline
RSC $\leq$ 50 & 0.25 & DEC $\leq$ 30 & 2\\
50 $<$ RSC $\leq$ 80 & 0.5 & 30 $<$ DEC $\leq$ 100 & 1.5\\
80 $<$ RSC $\leq$ 130 & 1 & 100 $<$ DEC $\leq$ 200 & 1\\\hline
\end{tabular}
\end{table}

\textbf{SST}: This comprises the time required to observe the number of agents and junctions as well as the relative speed coefficient. In this contribution, the 7 s TORTB that was previously concluded as appropriate for S1 and S3 is utilized to calculate the coefficients of number of agents $C_{NOA}$ and junctions $C_{NOJ}$ respectively.

\textbf{NOA}: The value of this coefficient is 1.9 per traffic agent which is obtained by substitution of S1 into equations (\ref{eqtib}) and (\ref{eqSST}). In other words, assuming that 7 s is the upper time bound for the slowest stimulus response time, least driving experience, reading NDRT, in the first drive, the coefficient $C_{NOA}$ by substitution into equations (\ref{eqtib}) and (\ref{eqSST}) is expressed as  

\begin{flalign}
\footnotesize
TORTB & = 0.3+2+2\times C_{NOA} + 0\times C_{NOJ}+1=7\\
C_{NOA} & = \frac{7 - 3.3}{2} = 1.85~s \approx 1.9~s.
\end{flalign}

This is a constant that can be substituted into equations (\ref{eqtib}) and (\ref{eqSST}) for another scenario.

\textbf{NOJ}: The value of this coefficient is 0.2 per number of junctions which was obtained by substitution of S3 variables into equations (\ref{eqtib}) and (\ref{eqSST}) as well as the value of $C_{NOA}$. In other words, assuming that 7 s is the upper time bound for the slowest stimulus response time, least driving experience, reading task, in the first drive, the coefficient $C_{NOJ}$ by substitution into equations (\ref{eqtib}) and (\ref{eqSST}) together with the previously computed value of $C_{NOA}$ is expressed as

\begin{flalign}
\footnotesize
TORTB&=0.3+ 2 + 2 \times 1.85 + 3 \times C_{NOJ} + 0.5 = 7 \\
C_{NOJ} & = \frac{7 - 6.5}{3} = 0.16~s \approx 0.2~s.
\end{flalign}

\textbf{RSC}: This is associated with the effort required to control the ego vehicle due to the current driving speed. This is relative to the nature of the cause of the critical situation which could be stationary (e.g. a junction) or moving (e.g a car). In this contribution, the assigned levels in Table~\ref{tb:Estimate_SRT_DEC_RSC} are substituted into Equations~\ref{eqtib} and \ref{eqSST} based on the characteristics of S1 and S3. In other words, the effect of increased speed is obvious between S1 and S3 given that the two scenarios have the same NOA but differ in the NOJ and ego vehicle speed. Furthermore given that~\cite{Hayashi:2021}, concluded that the time required by different drivers to observe another vehicle during takeover ranges between 0.8 s and 1.8 s, the obtained value of C$_{NOA}$ = 1.9 correlates. Moreover, substituting 1.8 s for C$_{NOA}$ in the context of S1 would result in a value for RSC that is not valid for substitution in the context of S3 given the difference in the ego vehicle speed of the two scenarios. Therefore, it is safe to assume that the given categorization of RSC are sufficient for substitution into the proposed formula.


\textbf{NDRTC}: Previous studies have concluded that TOT is significantly higher with NDRTs that involve the use of handheld devices compared to NDRTs that require handsfree devices as utilized in this contribution~\cite{YOON2019620,Nakajima:2017}. The TOT difference between handheld and handsfree NDRTs is 2.73 s~\cite{Nakajima:2017}. Accordingly, NDRTC = 0 s and 2.73 s for handsfree and handheld devices respectively.

\textbf{OC}: Takeovers may not be a frequent occurrence such that the first takeover for a scenario after a while or in each day has been shown to require more time \cite{Hergeth:2017,Wang:2019}. In this contribution, the same TORTB results in improved performance in relation to max. acc. in the third drive compared to the second. This learning effect indicates that varying the TORTB between repeated drives is still required. Accordingly, less time is required for the second and subsequent drives. Thus, the value of OC is taken as 0 s for the first drive and approximately 0.4 s for the subsequent drives. This is obtained by multiplying the ordinal effect size (partial $\eta^{2}$) 0.053 by the upper bound TORTB of 7 s for scenarios S1 and S3.

\textbf{Example of TORTB estimation:} The time required by a driver with SRT = 0.2 s and driving experience (80 km/wk) = 1.5 s using the aforementioned definitions and Table~\ref{tb:Estimate_SRT_DEC_RSC} for different variables are displayed in Table~\ref{tb:TORTB_ex}.


\begin{table*}[t]
\centering
\caption{Examples of TORTB estimation using different scenarios}~\label{tb:TORTB_ex}
\small
\begin{tabular}{p{0.45cm}p{0.45cm}p{2.0cm}p{1.05cm}p{0.9cm}p{1.6cm}p{1.6cm}}\hline
\multicolumn{3}{C{2.9cm}}{Scenario variables} & \multicolumn{3}{C{3.55cm}}{TOR variables} & TORTB [s]\\\cline{1-6}
NOA & NOJ & RS [km/hr] & RSC [s] & OC [s] & NDRTC [s]	& \\\hline
1 & 0 & 80 – 0 = 80 & 0.5 &  0  & 0 & 4.1 \\
2 & 0 & 130 – 0 = 130 & 1 &  0  & 0 & 6.5 \\
1 & 0 & 80 – 50 = 30 & 0.25 & 0  & 2.7 & 6.55 \\
2 & 0 & 130 – 50 = 80 & 0.5 & 0  & 2.7  & 8.7 \\
0 & 1 & 50 – 0 = 50 & 0.25 & 0.4  &  0 & 1.75 \\
0 & 1 & 100 – 0 = 100 & 1 & 0.4  & 0  & 2.5 \\\hline

\end{tabular}
\end{table*}

Given that the TOT often does not include the time the driver requires to complete the maneuver for the takeover, it would not have been appropriate to perform supervised learning and prediction using the TOT. In reality, some drivers will take over in less time than the estimated TORTB if they are highly experienced. However, 7 s has been confirmed to be optimal for both performance and SA for S1 and S3 given the scenarios' characteristics and the range of studied participants including the least experienced and slowest ones. Hence, having obtained the coefficients using the least experienced and slowest participant values, the formula is also adaptive to the more experienced and fast ones.


\section{Summary, conclusion, limitation, and outlook}
\subsection{Summary and conclusion}

This contribution integrates advance estimation of takeover time budget based on driver behavior. Variation of takeover time budgets that were defined based on average values obtained from a previous study for specific scenarios were evaluated for suitability. Together with scenario complexity, secondary task demand, and individual variables from previous studies and results from this contribution, a quantitative approach for budgeting time for takeover is proposed. The method enables an ADS to more precisely budget takeover time budget and determine in advance whether a driver will be able to successfully takeover in the available time. If the time will not be sufficient, additional measures such as shared precision control together with increased dynamic stability can be applied to ensure driver safety.

Furthermore, the warning imagery results in increased TOT although with reduced accidents. Thus if a suitably displayed information via an interface is generated, the time demand should be factored into the takeover time budget.

\subsection{Limitation and outlook}

Conditionally automated vehicles are not yet commercially available. The studied takeover scenarios are possibly less complex compared to those that frequently occur in reality. For example country roads include buildings, trees, and  parked cars that may additionally obscure a driver's view and require more takeover time budget. In addition, the participants were mostly alert and were not tired or fatigued. Moreso, the participants where young people who may utilize less time to takeover compared to elderly people. As an initial model it does not consider the spatial distribution of the agents in the scenarios. However, this contribution presents a systematic approach to estimate required takeover time budget and to extrapolate to more complex scenarios. 

The proposed quantitative approach for computing takeover time budget for drivers can be extended to integrate more variables (e.g., fatigue time demand) and be validated online for more complex scenarios. Specifically, the stimulus response time of each driver will be measured and integrated to vary takeover time budgets in the future. In addition, a suitable interface could reduce TOT and consequently takeover time budget but the formula makes it possible to estimate the time required without the interference of an interface.


\bibliographystyle{IEEEtran}
\bibliography{IEEEfull,IEEEabrv,mybibfile}

\vspace*{-9.5\baselineskip}

\begin{IEEEbiography}[{\includegraphics[width=1in,height=1.25in,clip,keepaspectratio]{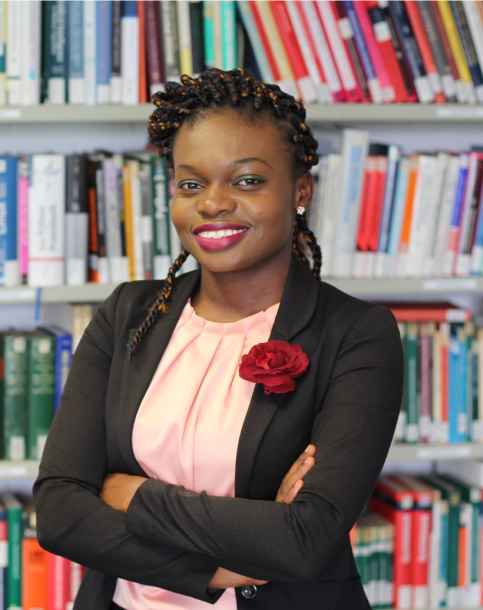}}]{Foghor~Tanshi}~(Member, IEEE) received a Dr.-Ing. degree in Mechanical Engineering from the University of Duisburg-Essen, Germany in 2021. Her current research interests include driver assistance systems and intelligent vehicles, traffic analysis, intelligent transportation systems, human-machine interfaces and interaction, human-machine cooperation, and cognitive systems.
\end{IEEEbiography}

\vspace*{-11.5\baselineskip}

\begin{IEEEbiography}[{\includegraphics[width=1in,height=1.25in,clip,keepaspectratio]{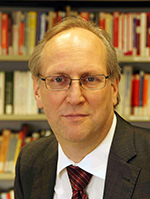}}]	{Dirk~S{\"o}ffker}~(M’10) (Member, IEEE) received a Dr.-Ing. degree
	in Mechanical Engineering and a Habilitation
	degree in Automatic Control/Safety Engineering from
	the University of Wuppertal, Wuppertal, Germany,
	in 1995 and 2001, respectively. Since 2001, he has
	been at the Chair of Dynamics and Control at the
	University of Duisburg-Essen, Germany. His current
	research interests include elastic mechanical structures, modern methods of control theory, human
	interaction with safe technical systems, safety and reliability control engineering of technical systems, and cognitive technical systems.
\end{IEEEbiography}

\EOD

\end{document}